\documentclass[aps,pre,showpacs,raggedbottom,nobalancelastpage,amssymb,twocolumn,groupedaddress]{revtex4}
\usepackage{graphicx}
\usepackage{amsmath}
\usepackage{amsfonts}
\usepackage{amssymb}
\usepackage{epsfig,libertine}
\usepackage[libertine]{newtxmath}
\usepackage{color}
\usepackage{dsfont, hyperref, xcolor}
\usepackage{comment}
\usepackage{enumerate}

\newcommand{\abs}[1]{\left\vert#1\right\vert}

\newcommand{\Tr}[1]{\text{Tr}\left\{#1\right\}}

\newcommand{\bra}[1]{\langle#1\vert}
\newcommand{\ket}[1]{\vert#1\rangle}

\begin{document}

\title{Fluctuation theorems and thermodynamic uncertainty relations}

\author{Gianluca~Francica}
\email{gianluca.francica@gmail.com}

\date{\today}

\begin{abstract}
Fluctuation theorems are fundamental results in non-equilibrium thermodynamics. Considering the fluctuation theorem with respect to the entropy production and an observable, we derive a new thermodynamic uncertainty relation which also applies to non-cyclic and time-reversal non-symmetric protocols. Furthermore, we investigate the relation between the thermodynamic uncertainty relation and the correlation between the entropy and the observable.
\end{abstract}

\maketitle
\section{Introduction}
In the last decades the non-equilibrium thermodynamics has received a great attention~\cite{kosloff13,Vinjanampathy16,bookthermo18}. Among the several results achieved in this field, the fluctuation theorems (see, e.g., Ref.~\cite{campisi11} for a review) undoubtedly are of fundamental importance. For instance, these theorems lead to the Jarzynski equality~\cite{Jarzynski97} and the second law of thermodynamics. Recently, it is shown how, for a cyclic and time-reversal symmetric protocol, these theorems imply the so-called thermodynamic uncertainty relations~\cite{Hasegawa19,Timpanaro19}, i.e. they give a lower bound of fluctuations of observables, which involves the entropy produced in the process. Conversely, for the case of general fluctuation theorems, hysteretic thermodynamic uncertainty relations~\cite{Proesmans19,potts19} and lower bounds for the entropy produced in the forward process~\cite{vo20,campisi21} are derived.

In this paper, we consider a more general fluctuation relation and derive a new thermodynamic uncertainty relation which is tighter than the one of Ref.~\cite{Proesmans19,potts19} and reduces to the one of Ref.~\cite{Timpanaro19} for a cyclic and time-reversal symmetric protocol. Furthermore, we investigate how it depends on the correlation between the entropy and the observable.

\section{Thermodynamic uncertainty relation}
We consider a generic system that starts in a thermal equilibrium state, such that for the entropy production $\sigma$ and the observable $\phi$ we have the multivariate fluctuation theorem
\begin{equation}\label{fluc.theo.}
\frac{p(\sigma,\phi)}{\tilde p(-\sigma,-\phi)} = e^\sigma\,,
\end{equation}
where $p(\sigma,\phi)$ and $\tilde p(\sigma,\phi)$ are the probability distribution functions for the joint occurrence of $\sigma$ and $\phi$ in the forward and backward process, respectively. In Ref.~\cite{Timpanaro19} a cyclic and time-reversal symmetric protocol is considered, for which $\tilde p(\sigma,\phi)=p(\sigma,\phi)$, and it is shown that in this case the Eq.~\eqref{fluc.theo.} implies
\begin{equation}\label{bound0}
\frac{\langle \phi \rangle^2}{\langle \phi^2 \rangle} \leq \left\langle \tanh^2\left(\frac{\sigma}{2}\right) \right\rangle\,,
\end{equation}
from which it follows the thermodynamic uncertainty relation
\begin{equation}\label{bound01}
\frac{\langle \phi^2 \rangle}{\langle \phi \rangle^2} \geq \frac{1}{f^2(\langle \sigma \rangle)}\,,
\end{equation}
where the function $f$ is the inverse of $ h(x)=2 x \tanh^{-1} x$ and the averages $\langle \cdots \rangle$ are calculated with respect to $p(\sigma,\phi)$, for instance $\langle \phi \rangle = \int \phi p(\sigma,\phi) d\sigma d\phi$.

Here, we consider the general case (i.e. we can have $\tilde p(\sigma,\phi)\neq p(\sigma,\phi)$), for which we have that the fluctuation theorem of Eq.~\eqref{fluc.theo.} implies
\begin{equation}\label{bound}
\frac{(\langle \phi\rangle + \langle \phi\rangle_R)^2}{\langle \phi^2\rangle + \langle \phi^2\rangle_R}\leq \left\langle \tanh^2 \left(\frac{\sigma}{2}\right)\right\rangle+\left\langle \tanh^2 \left(\frac{\sigma}{2}\right)\right\rangle_R\,,
\end{equation}
where the averages $\langle \cdots \rangle_R$ are calculated with respect to $\tilde p(\sigma,\phi)$.
We recall that an inequality equivalent to Eq.~\eqref{bound} has already been derived in Ref.~\cite{potts19}, where it is proved that from it follows the hysteretic thermodynamic uncertainty relation of Ref.~\cite{Proesmans19,potts19}
\begin{equation}\label{bound_known}
\frac{2(\langle \phi^2 \rangle+\langle \phi^2 \rangle_R)}{(\langle \phi \rangle+\langle \phi \rangle_R)^2} \geq \coth \left(\frac{\langle \sigma \rangle+\langle \sigma \rangle_R}{4}\right)\,.
\end{equation}
Here, we prove that from Eq.~\eqref{bound} follows the thermodynamic uncertainty relation
\begin{equation}\label{bound1}
\frac{2(\langle \phi^2 \rangle+\langle \phi^2 \rangle_R)}{(\langle \phi \rangle+\langle \phi \rangle_R)^2} \geq \frac{1}{f^2\left(\frac{\langle \sigma \rangle+\langle \sigma \rangle_R}{2}\right)}\,.
\end{equation}
Of course Eq.~\eqref{bound} and $\eqref{bound1}$ reduce to Eq.~\eqref{bound0} and \eqref{bound01} when $\tilde p(\sigma,\phi) = p(\sigma,\phi) $. Thus, the bound is saturated by the minimal distribution of Ref.~\cite{Timpanaro19}
\begin{eqnarray}
\nonumber p_{min}(\sigma,\phi) &=& \frac{1}{2\cosh(a/2)}\big( e^{a/2}\delta(\sigma-a)\delta(\phi-b)\\
 && + e^{-a/2}\delta(\sigma+a)\delta(\phi+b)\big)\,.
\end{eqnarray}
In particular, we have the relation (for $x\geq 0$)
\begin{equation}\label{inequalities}
\frac{2}{x}+1 \geq \frac{1}{f^2(x)} \geq \coth\left(\frac{x}{2}\right)\,,
\end{equation}
thus, the bound is smaller than $4/(\langle \sigma \rangle+\langle \sigma \rangle_R) +1$ and tighter than the one of Eq.~\eqref{bound_known}, derived in Ref.~\cite{Proesmans19,potts19}.

We note that, by defining the arithmetic mean $\langle \cdots\rangle_M = (\langle \cdots\rangle +\langle \cdots\rangle_R )/2$, i.e. the average with respect to $p_M(\sigma,\phi)=(p(\sigma,\phi)+\tilde p(\sigma,\phi))/2$, the Eq.~\eqref{bound1} reads
\begin{equation}\label{bound12}
\frac{\langle \phi^2 \rangle_M}{\langle \phi \rangle^2_M} \geq \frac{1}{f^2\left(\langle \sigma \rangle_M\right)}\,,
\end{equation}
which is our main result. It is worth noting that the Eq.~\eqref{bound12} is in agreement with Eq.~\eqref{bound01}, since the distribution function $p_M(\sigma,\phi)$ is such that $p_M(\sigma,\phi)/p_M(-\sigma,-\phi)=e^\sigma$.

We observe that how far the fluctuation is from the bound depends on the degree of correlation between the variables $\sigma$ and $\phi$, which can be characterized by the Pearson correlation coefficient defined as
\begin{equation}
\rho_{\sigma,\phi} =  \frac{\langle \sigma \phi \rangle - \langle \sigma \rangle\langle \phi \rangle}{\sqrt{(\langle \sigma^2\rangle-\langle \sigma \rangle^2)(\langle \phi^2\rangle-\langle \phi \rangle^2)}}\,.
\end{equation}
If there is no correlation we have the probability distribution $p(\sigma,\phi)=p(\sigma)p(\phi)$. Then, from the fluctuation theorem we have $\tilde p(\sigma,\phi)=e^{-\sigma}p(-\sigma) p(-\phi)$, such that $\langle \phi \rangle_R = -\langle \phi \rangle$ and $\langle \phi^2 \rangle_M/\langle \phi \rangle_M^2$ tends to infinity. Thus, we expect that the fluctuation is very far from the bound as the two variables are weakly correlated. On the other hand, the bound is saturated for the minimal distribution. In this case, it is easy to see that $\rho_{\sigma,\phi} = 1$ and there is a perfect correlation.
For completeness, we give the proofs of Eq.~\eqref{bound},\eqref{bound1} and \eqref{inequalities}.

{\it Proof of Eq.~\eqref{bound} --} We start by noting that for an arbitrary function $F(\sigma,\phi)$ we have
\begin{eqnarray}
\langle F(\sigma,\phi)\rangle &=& \int F(\sigma,\phi) p(\sigma,\phi) d\sigma d\phi\\
 &=& \int F(\sigma,\phi) e^\sigma \tilde p(-\sigma,-\phi) d\sigma d\phi\,,
\end{eqnarray}
where we have used the fluctuation theorem. Thus, by performing the substitutions $\sigma \to - \sigma$ and $\phi \to - \phi$ we get
\begin{equation}\label{rel_fund}
\langle F(\sigma,\phi)\rangle = \langle F(-\sigma,-\phi) e^{-\sigma}\rangle_R\,.
\end{equation}
In particular, from this relation we have that
\begin{equation}\label{rel x}
\langle x^n \rangle = (-1)^n \langle x^n e^{-\sigma} \rangle_R
\end{equation}
for $x=\sigma,\phi$, thus
\begin{equation}
(\langle \phi\rangle + \langle \phi\rangle_R)^2 = \langle \phi(1-e^{-\sigma})\rangle_R^2 = \left\langle \phi \sqrt{1+e^{-\sigma}}\frac{1-e^{-\sigma}}{\sqrt{1+e^{-\sigma}}}\right\rangle_R^2
\end{equation}
and by using the Cauchy-Schwartz inequality we have
\begin{equation}
(\langle \phi\rangle + \langle \phi\rangle_R)^2 \leq \langle \phi^2(1+e^{-\sigma})\rangle_R \left\langle \frac{(1-e^{-\sigma})^2}{1+e^{-\sigma}}\right\rangle_R\,.
\end{equation}
Since from Eq.~\eqref{rel x} it results that
\begin{equation}
\langle \phi^2\rangle + \langle \phi^2\rangle_R=\langle \phi^2(1+e^{-\sigma})\rangle_R\,,
\end{equation}
we obtain
\begin{equation}\label{rel int}
\frac{(\langle \phi\rangle + \langle \phi\rangle_R)^2}{\langle \phi^2\rangle + \langle \phi^2\rangle_R}\leq \left\langle \tanh^2 \left(\frac{\sigma}{2}\right)e^{-\sigma}\right\rangle_R+\left\langle \tanh^2 \left(\frac{\sigma}{2}\right)\right\rangle_R\,.
\end{equation}
By noting that the function $\tanh^2(\sigma/2)$ is even and by applying the relation in Eq.~\eqref{rel_fund} we get
\begin{equation}
\left\langle \tanh^2\left(\frac{\sigma}{2}\right) e^{-\sigma} \right\rangle_R = \left\langle \tanh^2\left(\frac{\sigma}{2}\right) \right \rangle\,,
\end{equation}
and thus from Eq.~\eqref{rel int} it follows the Eq.~\eqref{bound}.

{\it Proof of Eq.~\eqref{bound1}--} We define the functions $r(x)=\tanh^2(x/2)$, $k(x)=x \tanh(x/2)$, $g$ as a right inverse function of $k$ and $w$ as the composition of the functions $r$ and $g$, $w(x)=(r\circ g)(x)=r(g(x))$, which is concave, i.e. $w''< 0$. Since $g$ is a right inverse function of $k$ and $r$ is an even function, we have that $r(x)=r(g(k(x)))=w(k(x))$, where we have used the definition of $w$. Then, by using the Jensen's inequality, we have that $\langle r(\sigma) \rangle = \langle w(k(\sigma)) \rangle\leq w(\langle k(\sigma) \rangle)$, such that from Eq.~\eqref{bound} we have
\begin{equation}
\frac{(\langle \phi\rangle + \langle \phi\rangle_R)^2}{2(\langle \phi^2\rangle + \langle \phi^2\rangle_R)}\leq \frac{1}{2}w(\langle k(\sigma)\rangle) +\frac{1}{2}w(\langle k(\sigma) \rangle_R)\,.
\end{equation}
Since $w$ is concave, we have
\begin{equation}\label{rel int 2}
\frac{(\langle \phi\rangle + \langle \phi\rangle_R)^2}{2(\langle \phi^2\rangle + \langle \phi^2\rangle_R)}\leq w\left(\frac{\langle k(\sigma) \rangle+\langle k(\sigma) \rangle_R }{2}\right)\,.
\end{equation}
By noting that the function $k$ is even and by applying the relation in Eq.~\eqref{rel_fund} we get
\begin{equation}
\langle k(\sigma) \rangle+\langle k(\sigma) \rangle_R = \langle \sigma \tanh\left(\frac{\sigma}{2}\right)(1+e^{-\sigma}) \rangle_R = \langle \sigma(1-e^{-\sigma}) \rangle_R
\end{equation}
and from Eq.~\eqref{rel x} we get
\begin{equation}
\langle k(\sigma) \rangle+\langle k(\sigma) \rangle_R = \langle \sigma \rangle +  \langle \sigma \rangle_R\,.
\end{equation}
Thus, from Eq.~\eqref{rel int 2} we obtain
\begin{equation}
\frac{(\langle \phi\rangle + \langle \phi\rangle_R)^2}{2(\langle \phi^2\rangle + \langle \phi^2\rangle_R)}\leq w\left(\frac{\langle \sigma \rangle+\langle \sigma \rangle_R }{2}\right)\,.
\end{equation}
Taking the square root of both sides, applying then the increasing function $\tanh^{-1}$, multiplying them by $2$ and applying $k$, we get
\begin{equation}
h\left( \sqrt{\frac{(\langle \phi\rangle + \langle \phi\rangle_R)^2}{2(\langle \phi^2\rangle + \langle \phi^2\rangle_R)}}\right)\leq \frac{\langle \sigma \rangle+\langle \sigma \rangle_R }{2}\,,
\end{equation}
from which it follows Eq.~\eqref{bound1}.

{\it Proof of Eq.~\eqref{inequalities} --} Let's start considering the inequality $1/f^2(x) \geq \coth(x/2)$ with $x\geq 0$, which is equivalent to $\sqrt{\tanh(x/2)} \geq f(x)$. By defining $y=\sqrt{\tanh(x/2)}$, such that $0\leq y \leq 1$, and applying the function $h$ to both sides, we get the inequality $y\tanh^{-1} y \geq \tanh^{-1} y^2$. We have the series  $\tanh^{-1} z = \sum_{k=0}^\infty z^{2k+1}/(2k+1)$ for $\abs{z}\leq 1$, thus the inequality becomes $\sum_{k=0}^\infty y^{2k}/(2k+1) \geq \sum_{k=0}^\infty y^{4k}/(2k+1) $ which is true for $\abs{y} \leq 1$ since $y^{2k}\geq y^{4k}$ in this interval.
Nextly, we consider the inequality $2/x+1 \geq 1/f^2(x) $ with $x\geq 0$, which is equivalent to $f(x) \geq \sqrt{x/(2+x)}$.  By defining $y=\sqrt{x/(2+x)}$, such that $0\leq y \leq 1$, and applying the function $h$ to both sides, we get the inequality $y/(1-y^2) \geq \tanh^{-1} y$. We have the series  $1/(1-z) = \sum_{k=0}^\infty z^k$ for $\abs{z}\leq 1$, thus the inequality becomes $\sum_{k=0}^\infty y^{2k} \geq \sum_{k=0}^\infty y^{2k}/(2k+1) $ which is true for $\abs{y} \leq 1$.

\section{Physical example}
\begin{figure}
[h!]
\centering
\includegraphics[width=0.7\columnwidth]{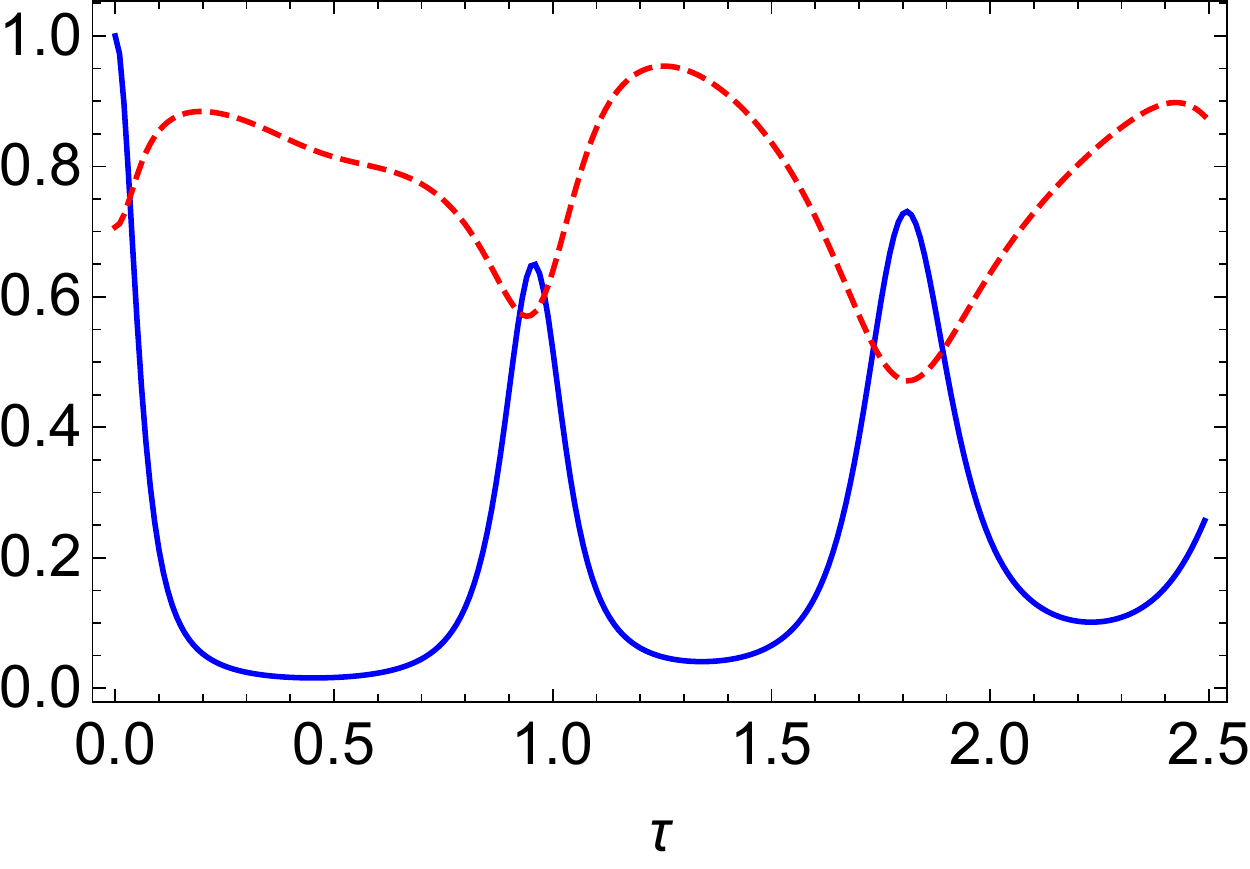}
\caption{ The plots of the normalized difference $\Delta_n$ (blue line)  and the correlation coefficient $\rho_{\sigma,\Delta E_A}$ (red dashed line) in function of the duration time $\tau$. In detail, the normalized difference $\Delta_n$ is obtained by dividing the difference $\Delta$ by its maximum value in the interval under consideration. We consider $\omega(t) = \omega_i + (\omega_f-\omega_i)t/\tau$ and $\lambda(t)=\lambda_0 \sin(\pi t/\tau)$ with $\omega_i=1$, $\omega_f=2$, $\lambda_0=4$ and $\beta=1$.
}
\label{fig:plot}
\end{figure}

We are interested to study how the difference $\Delta = \langle \phi^2\rangle_M/\langle \phi \rangle_M^2-1/f^2(\langle \sigma\rangle_M)$ is related to the correlation between $\sigma$ and $\phi$ characterized by the correlation coefficient $\rho_{\sigma,\phi}$. Thus, we proceed our investigation by considering a physical example that is a closed quantum system made of two parties $A$ and $B$. In the forward protocol, the time-dependent Hamiltonian $H(t)=H_A(t) + H_B(t) + H_{int}(t)$ generates the unitary time evolution operator $U_{t,0}$ defined by the Schr\"{o}dinger equation $i \dot U_{t,0}=H(t) U_{t,0}$ with initial condition $U_{0,0}=\mathds{1}$. In particular we consider the case in which the interaction $H_{int}(t)$ between the systems $A$ and $B$ is turned-off at the initial time $t=0$ and final time $t=\tau$. The initial state is the equilibrium state $\rho^G_i= e^{-\beta H(0)}/Z_i$ where $Z_i$ is the partition function $Z_i = \Tr{e^{-\beta H(0)}}$. By considering $\phi$ as the energy difference $\Delta E_A$ of the part $A$, we have the probability distribution function
\begin{eqnarray}
\nonumber p(\sigma,\Delta E_A) &=& \sum p^{(i)}_{m m'}P_{m m' n n'}\delta(\sigma-\sigma_{m m' n n'})\\
 && \times\delta(\Delta E_A- \epsilon^A_n(\tau)+\epsilon^A_m(0))\,,
\end{eqnarray}
where the initial populations are $p^{(i)}_{m m'}= \bra{\epsilon^A_m,\epsilon^B_{m'}(0)}\rho^G_i \ket{\epsilon^A_m,\epsilon^B_{m'}(0)}$, $\ket{\epsilon^X_m(t)}$ are eigenstates of $H_X(t)$ with eigenvalues $\epsilon^X_m(t)$, with $X=A,B$, and $\sigma_{m m' n n'}=\beta(\epsilon^A_n(\tau)+\epsilon^B_{n'}(\tau)-\epsilon^A_m(0)-\epsilon^B_{m'}(0)-\Delta F)$, where the difference of free energy is $\Delta F=-\ln(Z_f/Z_i)/\beta$ with $Z_f=\Tr{e^{-\beta H(\tau)}}$. The transition probability is given by $P_{m m' n n'}=\abs{\bra{\epsilon^A_n,\epsilon^B_{n'}(\tau)}U_{\tau,0} \ket{\epsilon^A_m,\epsilon^B_{m'}(0)}}^2$.
On the other hand, in the backward protocol, the initial state is $\rho^G_f= e^{-\beta H(\tau)}/Z_f$ and the time evolution from the initial to the final time is described by the operator $U_{\tau,0}^\dagger$. We have the probability distribution function
\begin{eqnarray}
\nonumber\tilde p(\sigma,\Delta E_A) &=& \sum p^{(f)}_{n n'} P_{m m' n n'}\delta(\sigma+\sigma_{m m' n n'})\\
 && \times\delta(\Delta E_A- \epsilon^A_m(0)+\epsilon^A_n(\tau))\,,
\end{eqnarray}
where $p^{(f)}_{n n'}=\bra{\epsilon^A_n,\epsilon^B_{n'}(\tau)}\rho^G_f \ket{\epsilon^A_n,\epsilon^B_{n'}(\tau)}$, such that the Tasaki-Crooks fluctuation theorem of Eq.~\eqref{fluc.theo.} is satisfied. In order to proceed with our study, we consider the particular case of two qubits with an Ising Hamiltonian, such that $H_X(t)=\omega(t) \sigma^X_z$, with $X=A,B$, and the interaction is $H_{int}(t) = \lambda(t) \sigma^A_x \otimes \sigma^B_x$, where $\sigma_x$, $\sigma_y$ and $\sigma_z$ are the Pauli matrices.
As shown in the Fig.~\ref{fig:plot}, the difference $\Delta$ is always positive and the bound of Eq.~\eqref{bound12} is satisfied. Furthermore, we have that the fluctuation deviates from the value of the bound when the correlation coefficient $\rho_{\sigma,\Delta E_A}$ decreases and so the correlation weakens.

\begin{figure}
[h!]
\centering
\includegraphics[width=0.7\columnwidth]{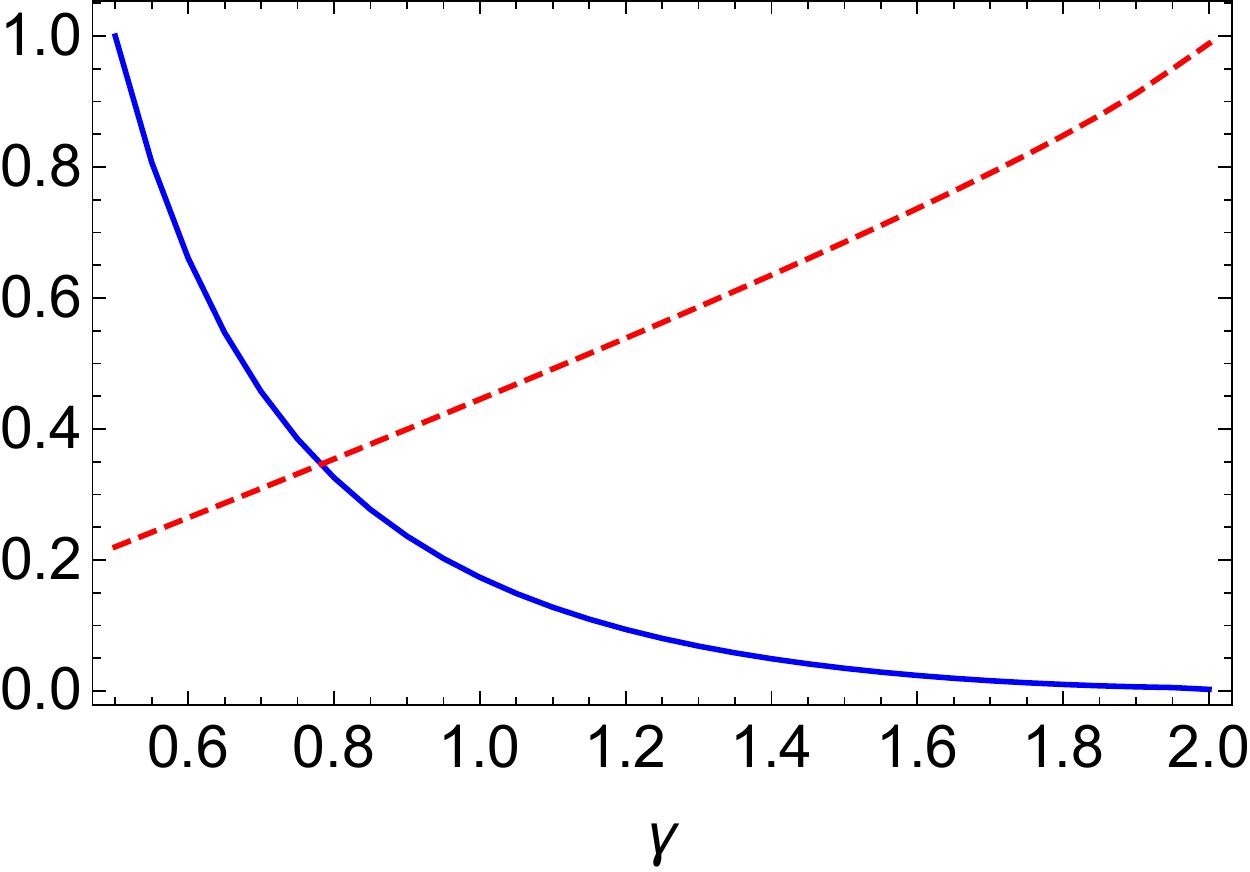}
\caption{ The plots of the normalized difference $\Delta_n$ (blue line)  and the correlation coefficient $\rho_{\sigma,\phi}$ (red dashed line) in function of the parameter $\gamma$.
}
\label{fig:plot2}
\end{figure}

Of course, how strong the relation between the difference $\Delta$ and the correlation is depends on the particular system under consideration.
In order to give another illustrative example, we note that given two non-negative functions $g_1(\sigma,\phi)$ and $g_2(\sigma,\phi)$, we can construct the probability
\begin{equation}
p(\sigma,\phi)= n (\hat g_1(\sigma,\phi) \theta(\sigma) + \hat g_2(-\sigma,-\phi) e^\sigma \theta(-\sigma))\,,
\end{equation}
such that satisfies the fluctuation relation. In detail, $n$ is a constant defined such that the probability $p(\sigma,\phi)$ is normalized, i.e. $\int p(\sigma,\phi) d\sigma d\phi = 1$, $\theta$ is the Heaviside step function, and the functions $\hat g_i$ are defined by
\begin{equation}
\hat g_i(\sigma,\phi) = \frac{g_i(\sigma,\phi)}{\int_{-\infty}^\infty d\phi \int_{0}^\infty d\sigma g_i(\sigma,\phi)(1-e^{-\sigma})}\,,
\end{equation}
with $i=1,2$, such that also $\tilde p(\sigma,\phi)=p(-\sigma,-\phi)e^\sigma$ is normalized. For simplicity, we consider $g_1(\sigma,\phi)=g_2(\sigma,\phi)=e^{-\sigma^2-\phi^2+\gamma \sigma \phi}$, where the parameter $\gamma$ determines the amount of correlation. In this simple case (in general for Gaussian functions) the difference $\Delta$ tends to decrease as the correlation tends to increase showing a very strong relation between the two (see Fig.~\ref{fig:plot2}).

\section{Conclusions}
In summary, we have investigated the relation between fluctuation theorems and thermodynamic uncertainty relations in the most general case of non-cyclic and time-reversal non-symmetric protocols, deriving a new lower bound of the observable fluctuation.
The bound is tighter than the one of Ref.~\cite{Proesmans19,potts19}, in detail the difference $d(x)=1/f^2(x)-\coth(x/2)$ gets its maximum value $1/3$ at $x=0$, is decreasing for $x\geq 0$ and $d(x)\to 0$ as $x\to \infty$, such that the two bounds  get closer as the average $\langle \sigma \rangle_M$ increases.
Furthermore, we have performed a study in order to clarify the relation between the thermodynamic uncertainty relation and the correlation between the two variables $\sigma$ and $\phi$. In particular, we find that the fluctuation tends to be far from the bound as the two variables are weakly correlated.

\end{document}